\documentclass[12pt,openbib]{article}
\usepackage{graphicx}                 
\usepackage{color}                    
\usepackage{hyperref}                 

\begin{document}

\title{Bending a slab of neural tissue}
\author{Partha P.Mitra\\Cold Spring Harbor Laboratory, NY 11724}
\maketitle

\begin{abstract}

In comparative and developmental neuroanatomy one encounters questions regarding the deformation of neural tissue under stress. The motivation of this note is an observation (Barbas {\it et al}\cite{barbas}) that at cortical folds or gyri, the layers of neural tissue show relative thickening or thinning of upper or deep layers. In general, the material properties of a slab of neural tissue are not known, and even if known, would probably lead to a difficult problem in elasticity theory. Here a simple argument is presented to show that bending an elastic slab should produce a relative thickening of the layers on the inside of the bend. The argument is based on the incompressibility of the material and should therefore be fairly robust. 
	
\end{abstract}

\newpage

We consider the bending of a slab of neural tissue. The material properties of such tissue are going to be complicated; it won't in general be a simple elastic medium\footnote{Note also that when subjected to large stresses the tissue will probably rupture. The arguments here have to be treated with a number of cautions discussed at the end of the manuscript}. One might expect the tissue to have both elastic properties, as well as some degree of fluidity (visco-elastic). However, on the whole one might expect the tissue to be more or less incompressible while retaining some elastic properties. This simple idea has some implications as to what happens when one bends a slab of such tissue. 

Consider a rectangular slab of tissue (see Fig.1) of length L and width W.


We won't consider the third dimension - the assumption is that this dimension is left unchanged by the bending (by applying suitable forces). 
In general, when this slab is bent, one will have to solve a complicated boundary value problem for the associated elasticity problem (which will not in general be in the regime of linear deformations). However, there is one special case which may be easily treated using symmetry, namely when this slab of tissue is bent into a circular shape (see figure 2). We assume this can be done in a reversible way without rupturing or fracturing the material.


 In this case, straight lines in the original rectangular slab can be expected to be bent into concentric circles or radial spokes. We show one such line in the figures. The line shown is parallel to the length of the slab, at a distance $Y$ from the lower edge of the slab, and is bent into the shape of a circle at radius $R(Y)$.
 
If we now assume that the material is incompressible, then the area bounded by the rectangle below this line must be equal to the area of the corresponding annulus. We therefore have $L Y = \pi [R^2(Y)- R_0^2]$. Solving this equation for the radius $R(Y)$, we have 

\begin{equation}
R(Y) = \sqrt{ {L Y \over \pi} + R_0^2}
\end{equation}

In general, $R(Y)-R_0$ will not be equal to $Y$, so there is an expansion or contraction of the corresponding thickness, given by

\begin{equation}
{R(Y)-R_0 \over Y} = {\sqrt{ {L Y\over \pi} + R_0^2} - R_0 \over Y}
\end{equation}

Note that there is one unknown parameter in the above, namely $R_0$. This radius is not determined if we simply use incompressibility. However, a homogeneous isotropic elastic is governed by two parameters, the compressibility and the shear moduli. In assuming incompressibility, we are in effect assuming that the compressibility modulus is much larger than the shear modulus. If we assume that the shear modulus is negligible compared to the compressibility modulus but nevertheless nonzero, this will determine the remaining parameter. Note however that the deformation is not small, and will not in general be in a linear elastic regime, so we have not carried out this last calculation. However, qualitative arguments may be used to show that $R_0$ has a value that does lead to the lower layers of the slab to be thickened, namely $\sqrt{ {L Y\over \pi} + R_0^2} - R_0 > Y$ for small enough $Y$. After some algebra, this condition can be re-written as 

\begin{equation}
{L - 2\pi R_0 \over \pi} > Y
\end{equation}

This condition can always be satisfied for small $Y$ if 
$L - 2\pi R_0 > 0$. This last condition asserts that the interior circumference of the annulus is less than the length of the slab. We can give a qualitative argument that this is the case by considering the shear energy on bending the slab. Consider a thin slice of the slab that is bent into a thin annulus. The greater the stretching or contraction of this annulus, compared to the length of the initial thin slice, the greater will be the amount of shear energy\footnote{It may seem odd to associate shear energy with a stretching or contraction. Note however that since the slab is being bent, the deformation is not isotropic and there will be some shear strains which will increase with the amount of stretching or compression of the annulus.}. Therefore, from energy minimization considerations, one would expect that there will be a line somewhere in the middle of the slab, that is neither expanded or contracted (so that at least this region of the slab does not contribute shear energy). This would be a line of no stretching. Since this line is in the middle of the slab, the slab will be thickened below this line and will be thinned above this line. 

The alternative, of course, is that the slab is entirely thickened or thinned - we are arguing that this will cost more shear energy than the configuration in which the slab has a thickened lower layer, and a thinned upper layer. 

The position of the non-stretched line will be given by the equation $L=2\pi R(Y)$. Assuming that we are dealing with a thin slab so that this line of no stretching is in the exact middle of the slab, we obtain the equation $L=2\pi R(W/2)$ which can be solved for $R_0$. For a thick slab, the exact value of $R_0$ will depend on nonlinear elastic parameters and will be difficult to estimate {\it a priori}. However, the arguments still show that the inner layers of the slab will be thickened when bent into an annulus. 

Note that we have bent the tissue into an annulus in order to use symmetry arguments. The contour lines will not be quite so nicely symmetrical if the slab is bent into a hairpin rather than an annulus. Nevertheless, one expects the qualitative arguments for the thickened inner layers to still hold. 

The arguments advanced above also tell us something about what happens to the total thickness of the slab when bent (as opposed to the relative thickness of layers). The thickness after bending is given by $R_1-R_0$, where $R_1$ is the outer radius of the annulus. Incompressibility translates to conservation of area, so that $\pi(R_1^2-R_0^2)=LW$. Therefore, the thickness of the annulus is given by 

\begin{equation}
R_1-R_0 = {LW \over \pi (R_1+R_0)} 
\end{equation}

Now if we assume, based on the argument given above about minimising shear energy, that there is a radius inside the annulus with circumference corresponding to the unstretched length, then $2\pi R_1>L>2\pi R_0$. Substituting this in the above equation we obtain, 

\begin{equation}
{2 R_1\over R_1+R_0} W >R_1-R_0 > {2 R_0 \over R_1+R_0} W
\end{equation}

This would allow both for an overall thinning or thickening of the whole slab, within the constraints provided by the equation given above\footnote{I am grateful to Dr Manajit Sengupta for pointing out an algebraic error in an earlier version of the manuscript in the derivation of the total thickness of the slab, leading to the incorrect conclusion that the slab would be always thickened overall}. 

Finally, a discussion is in order about the treatment of neural tissue as an elastic medium. First, note that most of the argument given here relies on incompressibility, a robust assumption that does not depend on the details of the material properties of the tissue and should hold under somewhat more general circumstances. Secondly, one should exercise care in applying these arguments to a developmental or evolutionary context. In the evolutionary context, where one compares the brains of different species with different extents of cortical folding, investigators have invoked a fictitious tension to associate a cost with axonal length. This tension is fictitious since the argument involves selective pressures on axonal length (due, for example, to metabolic costs associated with longer axons, or increased propagation delays). This does not necessarily  relate to the material properties of neural tissue\footnote{Note that this does not of course {\it preclude} the existence of physical tensions in neural tissue in the brain of a given species}. 

In the developmental context, one may expect physical elastic stresses of the sort discussed in this note to be of somewhat more relevance. Even here one ought to be cautious. In the usual material physics context, elastic stresses originate in the deformation of atomic bonds of a homogeneous material. Neural tissue is microscopically heterogeneous, and is composed of living neurons which are plastic and are potentially able to physically grow or shrink in response to mechanical stresses over developmental timescales, unlike the material physics example. Therefore, over such long timescales, arguments based on elasticity should be used with caution.

\end{document}